\documentclass[conference]{IEEEtran}
\usepackage{cite}
\def\BibTeX{{\rm B\kern-.05em{\sc i\kern-.025em b}\kern-.08em
    T\kern-.1667em\lower.7ex\hbox{E}\kern-.125emX}}
\usepackage{csquotes}

\usepackage{array}

\usepackage{graphicx}

\begin{document}
%
\title{Can Requirements Engineering Support Explainable Artificial Intelligence? Towards a User-Centric Approach for Explainability Requirements
}

\author{\IEEEauthorblockN{Umm-E- Habiba}
\IEEEauthorblockA{Institute of Software Engineering\\
University of Stuttgart\\
Stuttgart, Germany\\
Email: umm-e-habiba@iste.uni-stuttgart.de}
\and
\IEEEauthorblockN{Justus Bogner}
\IEEEauthorblockA{Institute of Software Engineering\\
University of Stuttgart\\
Stuttgart, Germany\\
justus.bogner@iste.uni-stuttgart.de}
\and
\IEEEauthorblockN{Stefan Wagner}
\IEEEauthorblockA{Institute of Software Engineering\\
University of Stuttgart\\
Stuttgart, Germany\\
stefan.wagner@iste.uni-stuttgart.de}}


%


\maketitle

\begin{abstract}
With the recent proliferation of artificial intelligence systems, there has been a surge in the demand for explainability of these systems.
Explanations help to reduce system opacity, support transparency, and increase stakeholder trust. In this position paper, we discuss synergies between requirements engineering (RE) and Explainable AI (XAI). We highlight challenges in the field of XAI, and propose a framework and research directions on how RE practices can help to mitigate these challenges.
\end{abstract}

%
\IEEEpeerreviewmaketitle

\section{Introduction}
In the last decade, Artificial Intelligence (AI) systems, especially based on Machine Learning (ML), have achieved remarkable feats in many areas that were previously computationally impossible \cite{lecun2015deep}.  AI is now prevalent in our professional and personal lives. Progress has also been made in the fields of medicine, automated vehicles, bioinformatics, and recommender systems. Despite these advancements, trust issues and disillusionment are emerging in several areas.
One of the most common problems is the lack of transparency in these systems, as the black-box nature of many ML models collides with the human desire to understand the system and its output \cite{yang2021does}. This is especially concerning when these systems are used in contexts where they have a major impact on human lives, such as medical diagnostic systems or financial and legal matters.

Software systems' growing autonomy and complexity make it harder for software engineers and domain experts to comprehend them \cite{darpa2016broad}, especially if they are composed of ML components \cite{lipton2018mythos}. This leads to a need for explainable systems. Explanations help to understand system decisions, which increases confidence in a system and improves trustworthiness. Additionally, it also justifies  actions taken, increases usability, aids in uncovering causes of mistakes, and reduces the potential of human error because humans cannot make informed decisions without access to the system rationale for its recommendation. 
An appropriate explanation for unexpected or inaccurate system behavior might also help to determine problems, for instance, the misinterpretation of a requirement or a mistake in the system design.

Moreover, a lack of explainability not only causes ethical, social, and legal issues \cite{mittelstadt2016ethics, floridi2018ai4people}. It also exacerbates distrust, reduces user acceptability \cite{lahijanian2016social, ye1995impact} and stifles the adoption of innovative technology. 
In this regard, the Ethics Guidelines for Trustworthy AI \cite{ai2019high} have been proposed by the High-Level Expert Group on Artificial Intelligence (AI HLEG).
They highlighted transparency as a key requirement for trustworthy AI, which is further comprised of traceability, explainability, and communication. Depending on the context of relevant stakeholders, these guidelines emphasize a proper explanation of the decision-making process of AI systems. Suitable development techniques are required to provide a certain level of explainability. The development process should incorporate explainability to make it explicit. However, the concrete scope for \emph{explainability} is still unclear, and many researchers are actively exploring different courses of action to support explainability.
While approaches such as LIME \cite{ribeiro2016should} and Shapley values \cite{hall2019systematic} are based on mathematical methods, the Requirements Engineering (RE) community is also exploring different approaches \cite{wolf2019explainability, cirqueira2020scenario} to support explainability, although it has already been identified as a key non-functional requirement to support transparency and reduce opacity. Current research is either inclined towards model explainability, which particularly focuses on ML engineers, or user experience, which addresses user needs for explainability and how users react to certain explainability techniques. A problem arises when it comes to the interaction between ML engineers and multiple stakeholders, as they have different expectations and views of explainability. However, there is a lack of studies that investigate the gap between the understanding of ML engineers and end-users regarding explainability. We believe RE practices can help to address this challenge.

\textbf{Position and Contribution:}
In this paper, we provide an overview of the state-of-the-art in Explainable AI (XAI). Our objective is to highlight current challenges in XAI from the perspective of requirements engineering and provide future directions on what could be options to mitigate these challenges.

\label{sec:RelatedWork}
\section{Explainability vs. Interpretability}
Terminology related to explainability in machine learning is not consistently used \cite{kohl2019explainability}, with terms such as \enquote{explainable AI} \cite{samek2017explainable}, \enquote{interpretability} \cite{doshi2017towards}, and \enquote{comprehensibility} \cite{freitas2014comprehensible}. They are all aimed at making the system more transparent, i.e., explaining the system’s inner working and making it more understandable.

In particular, explainability and interpretability appear frequently in the literature, and many researchers use them interchangeably \cite{carvalho2019machine} as no specific mathematical definitions for explainability or interpretability exist, nor have they been quantified by any metric \cite{adadi2018peeking}. While they refer to overlapping concepts, there is some work that attempts not only to define these two concepts but also related notions such as comprehensibility \cite{lipton2018mythos, doshi2017towards, gilpin2018explaining}.
Miller \cite{miller2019explanation} defines interpretability as \enquote{the degree to which a human can understand the cause of a decision}. Another definition comes from Doshi-Velez and Kim \cite{doshi2017towards}, who define interpretability as \enquote{the ability to explain or communicate in intelligible words to a human}. These definitions refer to interpretability as a property of how input and output are associated and how easily end-users can identify cause-and-effect relations.

In contrast, \enquote{explainability} deals with the internal logic of the system, i.e., how the ML system has been trained and how a particular output has been generated \cite{turek2018explainable}. 

Doshi-Velez and Kim \cite{doshi2017towards} define explanations as \enquote{the currency in which we exchange beliefs}. 
Rudin \cite{rudin2019stop} differentiates between interpretability and explainability via the concepts \enquote{inherently interpretable} and \enquote{post-hoc explanations}.
In \cite{doshi2017towards} and \cite{gilpin2018explaining}, the authors describe interpretability as a broader concept and argue that explainability is essential to support interpretability.

\begin{table*}[]
\centering
\caption{Interpretability vs Explainability in AI Literature}
\label{tab:ExpVsInter}
\begin{tabular}{lll}
\textbf{Papers} & \textbf{Interpretability} & \textbf{Explainability} \\ \hline
\multicolumn{1}{l}{\textbf{\cite{lipton2018mythos}}} & \multicolumn{1}{l}{Interpretability addresses \enquote{how does the model work?}} & \multicolumn{1}{l}{Explainability addresses \enquote{what else can the model tell me?}} \\ 
\multicolumn{1}{l}{\textbf{\cite{doshi2017towards}}} & \multicolumn{1}{l}{\begin{tabular}[l]{@{}l@{}}Interpretability is "the ability to explain or to\\ present something in understandable terms to a human"\end{tabular}} & \multicolumn{1}{l}{Explanations are "the currency in which we exchange beliefs"} \\ 
\multicolumn{1}{l}{\textbf{\cite{gilpin2018explaining}}} & \multicolumn{1}{l}{\begin{tabular}[l]{@{}l@{}}Mostly concerned with the intuitions behind the output of a model\\"to describe the internals of a
system in a way\\that is understandable to humans"\end{tabular}} & \multicolumn{1}{l}{\begin{tabular}[l]{@{}l@{}}Mostly concerned with the internal logic, i.e.,\\ "models that are able to summarize the reasons for neural network \\behavior, gain the trust of users, or produce
insights about \\ the causes of their decisions"\end{tabular}} \\ 
\multicolumn{1}{l}{\textbf{\cite{miller2019explanation}}} & \multicolumn{1}{l}{\begin{tabular}[l]{@{}l@{}}“The degree to which an observer can\\ understand the cause of a decision"\end{tabular}} & \multicolumn{1}{l}{\enquote{Explicitly explaining decisions to people}} \\ 
\multicolumn{1}{l}{\textbf{\cite{montavon2018methods}}} & \multicolumn{1}{l}{\begin{tabular}[l]{@{}l@{}}"An interpretation is the mapping of an abstract concept\\ into a domain that humans can make sense of"\end{tabular}} & \multicolumn{1}{l}{\begin{tabular}[l]{@{}l@{}}"An explanation is the collection of features of the interpretable\\ domain that have contributed to a given example to produce a decision"\end{tabular}} \\ \hline
\end{tabular}
\end{table*}

Table~\ref{tab:ExpVsInter} summarizes how different authors strived to define explainability and interpretability. However, we can see from this literature, there are no agreed definitions of explainability or interpretability. The AI research community is still working towards a unified position. To motivate research in this area, researchers are trying to understand and elicit different requirements, e.g., why these explanations and interpretations are needed, what their context is, and who the stakeholders in these explanations are.

To address end-user needs, research must consider their perspective of explanations. While the human-computer interaction (HCI) community is actively exploring user-centric XAI \cite{ferreira2021designer,abdul2018trends,wang2019designing}, a more extensive contribution is still required from the RE community to provide a systematic approach to enable software engineers to effectively incorporate explainability requirements in the development process. In the following section, we propose an alliance between RE and XAI.

\section{Synergies between RE and XAI}
\label{sec:synergies}

The interactions between the two fields RE and XAI are still evolving. A few studies discuss various interactions of RE and XAI. However, the major focus are concrete requirements for XAI. Mark et al.~\cite{hall2019systematic} proposed a 5-step methodology to understand explainable AI requirements. Their goal was to understand explanation requirements from different perspectives. Kohl et al.~\cite{kohl2019explainability} proposed a process for eliciting, specifying, and verifying explainability as a non-functional requirement. Chazette et al.~\cite{chazette2019end} also concentrated on a user perspective of explanations. Their work is more focused on software transparency and user opinion about embedded explanations in software. Another work \cite{chazette2020explainability} from the same authors explores the interaction between explainability and other non-functional requirements. They also describe potential trade-offs of incorporating explainability into the system. Similarly, Liao et al.~\cite{liao2020questioning} interviewed 20 user experience and design practitioners to identify gaps in current practices. They provided insights into the design space of XAI and a question bank that can be used to create user-centered XAI. Eiband et al.~\cite{eiband2018bringing} improved existing user interface guidelines by proposing a stage-based participatory process. Their approach is aimed more towards design management and user interface design.

Most of the presented work is focused on user-centered design and interactions of explainability with other non-functional requirement. There are studies centered on different stakeholder perspectives, but several challenges remain. We will outline them in the following section.

\section{Challenges}
\label{sec:challenges}
Current research in XAI is inclined towards model explainability and that these explanations are more focused on ML engineers rather than end-users of the system. Most existing explanation techniques and tools are only comprehensible by technical stakeholders with an ML background, like data scientists and ML engineers. Approaches that pay attention to the end-user are more rare. Although there are studies that attempted to work in the direction of requirements engineering and try to ask stakeholders about XAI requirements, many challenges remain. 

\paragraph{Absence of a mediator role}
Recently, several open-source toolkits \cite{marcinkevivcs2020interpretability} emerged for XAI. They provide explanations which are mostly comprehensible by ML engineers. Yet, these approaches are difficult to adapt for end-users, e.g., due to the gap between user needs and available technical capabilities to improve the effectiveness of explainability in AI products \cite{liao2020questioning}. Thus, the role of mediators arises, whose job it is to fill the gap between stakeholders. They have the experience to identify user needs and to convey them to the developers, which supports a better translation of end-user needs into technical aspects. In this sense, they serve as a bridge between users and ML engineers, keeping in mind the demands and constraints on both sides. Thus, explainable solutions for AI systems also need to provide explanations for the mediators, whose job is to bridge the gaps among stakeholders.

\paragraph{No coherent definition of explainability}
Usually, demands for explainability and how it can be accomplished are ambiguous~\cite{kohl2019explainability}. Several authors presented different facets of explainability, with no agreed on definition, which can, e.g., impact communication in AI projects.

\paragraph{Lack of stakeholder-centric approaches}
Current solutions for model interpretability do not describe or specify their intended users. Therefore, most of these solutions unwittingly wind up being more understandable to the people who build them, i.e., ML engineers. Alternately, the interpretable model built for the  end-user has issues such as simple visuals with general explanations that are often not useful for people in practice \cite{suresh2021beyond}. 

\paragraph{No common vocabulary for all ML stakeholders}
Many terms have recently been introduced in explainable AI. However, these terms appear interchangeably, which leads to confusion in this rapidly expanding field. Furthermore, certification authorities, researchers, end-user, domain experts, and ML engineers all use explainable AI with different expectations and understanding. This leads to more problems, as each stakeholder desires separate objectives to be satisfied by the term explainable AI \cite{brennen2020people}.

To address these challenges, we propose a general framework to provide guidance how these issues can be minimized and how explainability can be better supported by RE practices.
For each step, we are working on more detailed methods and guidelines to support practitioners.

\section{Proposed Framework}
\label{sec:proposedframework}

In this section, we i) draw a boundary between interpretability and explainability and ii) propose a framework to mitigate the challenges stated above.

As we have seen, these two terms have been used interchangeably, even though some authors draw a boundary between them. As a synthesis of these definitions, we differentiate between the two terms to make it explicit that  our framework supports explainability:

\begin{enumerate}
    \item \textit{Interpretability}: looking \textit{at} the model, i.e., the stakeholder can understand the cause-and-effect relationship between input and output. 
    \item \textit{Explainability}: looking \textit{inside} the model, i.e., the stakeholder can understand the inner workings of the model and how the model has produced a particular output.
\end{enumerate}

  \begin{figure*}[htb]
  	\includegraphics[width=\textwidth]{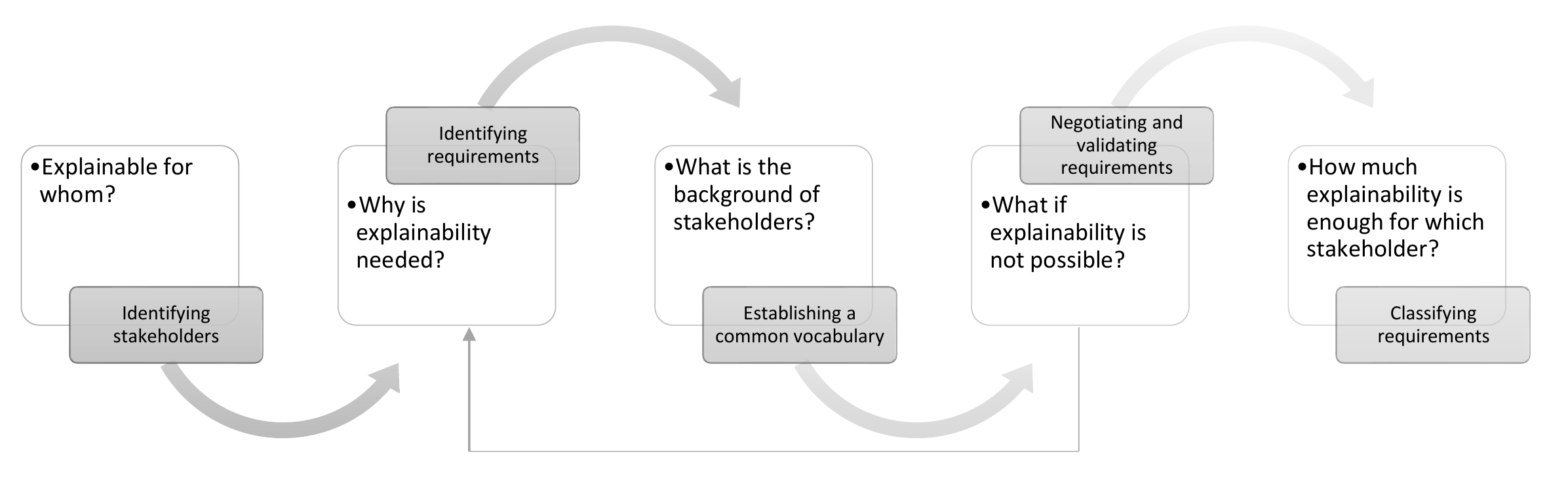}
  	\centering
  	\caption{Proposed Framework}
  	\label{fig:framework}
  	\vspace{-3mm}
  \end{figure*}

Activities described in our framework will be performed by the mediating role of the requirements engineer. This framework explicitly aims to identify stakeholders with the requirements of explainability. Furthermore, it will help to establish a shared understating of explainability among multiple stakeholders.
The framework comprises the following steps (Figure~\ref{fig:framework}):
\begin{enumerate}
    \item \textbf{Identifying stakeholders:} In this step we aim to address the question \enquote{explainable for whom?} We will identify relevant stakeholders for the system, i.e., end-users, ML engineers, and domain experts. We will also determine the stakeholders for explanations.
    \item \textbf{Identifying requirements:} We will elicit and document the concrete requirements for explainability, together with rationales like why explainability is required and what the impact of each explanation will be on the system. We will also describe what types of explanations will be provided. This will answer the question \enquote{why is explainability needed?}
    \item \textbf{Establishing a common vocabulary:} Using knowledge of end-users, domain experts, and ML engineers, we will establish a common vocabulary. We will introduce common terms and create a shared system understanding, which supports discussing explainability requirements.
    \item \textbf{Negotiating and validating requirements:} This is an iterative process. The identified explainability requirements will be validated in this step. This will answer questions like \enquote{what if explainability is not possible?} or \enquote{how much uncertainty is tolerable?} If there is any trade-off between the requirements or some are not feasible, they will be negotiated.
    \item \textbf{Classifying requirements:} In this step, requirements will be classified with respect to stakeholders. Who are the stakeholders for  particular explanations, and what do they want to achieve with these explanations? The main question to answer is \enquote{how much explainability is enough for which stakeholder?} This classification will also help to address each stakeholder requirements.
\end{enumerate}

\section{Conclusion and Future Work}
\label{sec:conclusion}
In this position paper, we discussed differences between interpretability and explainability, and pointed out existing challenges in the XAI space. Taking a requirements engineering perspective, we propose a work-in-progress framework to address the challenges, with emphasis on a user-centered approach to explainability requirements and highlight the potential of RE to support the explainability of AI systems.



\bibliographystyle{IEEEtran}
\bibliography{references}

\end{document}